\documentclass[12pt,preprint]{aastex}

\slugcomment{To appear in ApJ}

\shorttitle{Diffuse X-ray emission from NGC~2403 discovered with Chandra}
\shortauthors{Fraternali, Cappi, Sancisi, Oosterloo}

\begin{document}

\title{Diffuse X-ray emission from the spiral galaxy NGC~2403 discovered with Chandra}

\author{Filippo Fraternali}
\affil{Osservatorio Astronomico, via Ranzani 1, 40127, Bologna, Italy}
\email{ffratern@ira.bo.cnr.it}                                 
                
\author{Massimo Cappi}
\affil{Istituto TeSRE (CNR), via Gobetti 101, 40129, Bologna, Italy}
\email{mcappi@tesre.bo.cnr.it}

\author{Renzo Sancisi}
\affil{Osservatorio Astronomico, via Ranzani 1, 40127, Bologna, Italy} 
\affil{Kapteyn Astronomical Institute, University of Groningen, The
Netherlands}
\email{sancisi@bo.astro.it}                                    
                
\author{Tom Oosterloo}
\affil{ASTRON,  P.O. Box 2, 7990 AA, Dwingeloo, The Netherlands}
\email{oosterloo@nfra.nl}

\begin{abstract}
{
We have detected diffuse soft X-ray emission (0.4$-$1 keV) from the
disk of the spiral galaxy NGC~2403 with {\it Chandra}. 
This diffuse emission (with a total luminosity of 
2.1 $\times$ 10$^{38}$ erg
s$^{-1}$ and a gas temperature of 2$-$8 $\times$ 10$^{6}$ K) 
is well separated from the numerous bright point sources. 
NGC~2403 is a luminous spiral galaxy with a high rate of star formation. }
Recent H~{\small I} observations have revealed 
an extended H~{\small I} halo with anomalous velocities and a
general inflow towards the central regions of the galaxy. 
This result and the present detection of a diffuse, hot X-ray
emitting gas point at a very active disk-halo connection
and galactic fountain types of phenomena.
\end{abstract}


\section{Introduction}

Before the launch of {\it Chandra} the study of diffuse X-ray emission
from normal (non-starburst) spiral galaxies was seriously impeded
by lack of spatial resolution \citep{fab89}.
The point sources could not be well separated from the diffuse thermal gas.
The detection of diffuse emission from the disk of spiral galaxies was therefore
only possible for some nearby objects
(e.g.\ M~33: \citet{lon96}; M~101: \citet{sno95}; M~51: \citet{ehl95}).
Coronal (halo) emission was observed in some nearby
edge-on spiral galaxies like NGC~891 \citep{bre97} and NGC~4631 \citep{wan95, wan01}.
Also the diffuse component of the hot thermal plasma in the Milky Way has been
studied extensively (e.g. \citet{kan97, val98}).

Here we present {\it Chandra} observations of the spiral galaxy NGC~2403
that made it possible to unambiguously separate discrete sources from diffuse
emission. 
This galaxy is a nearby Sc spiral, morphologically similar to M~33,
viewed at an inclination angle of $\sim$~60$^{\circ}$.
It is well isolated on the sky and shows no signs of recent interactions.
H~{\small I} observations \citep{fra02} have revealed
a kinematically anomalous component of neutral gas. 
The H~{\small I} position-velocity (p-v) diagram along the major axis of this galaxy
(Figure~\ref{fig1}) shows systematic asymmetries in the
form of wings in the line profiles at lower velocities 
with respect to the rotation curve (white squares).
In the central part of the galaxy such wings extend up to 150 km s$^{-1}$
from the rotation curve.
This kinematical pattern is different from that expected for a thin cold disk
of H~{\small I} (see the model in the right panel of Figure~\ref{fig1}).
This anomalous gas was previously unknown.
It has not been detected before, mainly because of lack of sensitivity 
of the observations.

A detailed analysis of the H~{\small I} data of NGC~2403 has shown that 
the wings of H~{\small I} are produced by gas (the ``anomalous gas'')
located above the plane of the galaxy
and rotating $\sim$~20$-$50 km~s$^{-1}$ more slowly than the gas in the disk.
A similar slower rotation of the ``halo'' gas had already been observed in the
edge-on galaxy NGC~891 \citep{swa97}.
The anomalous gas in NGC~2403
extends out to $\sim$ 15 kpc from the centre of the galaxy and
has a total mass of about 3 $\times$ 10$^8$ 
$M_{\odot}$ ($\sim$~1/10 of the total H~{\small I} mass). 
The study of its velocity field has also revealed 
a probable large-scale radial inflow (10$-$20 km~s$^{-1}$) towards the centre of
the galaxy \citep{fra01}.

A possible interpretation is that of a {\it galactic fountain} \citep{sha76, bre80} and
the observed neutral gas may be the result of cooling of ionised gas blown up
from the disk into the halo.
The discovery of hot X-ray emitting gas from the disk of NGC~2403, reported here, 
supports this picture.
The anomalous H~{\small I} may be related to the hot X-ray gas.

\section{Observations}

NGC~2403 was observed with {\it Chandra} in April 2001 for a total time of
36~ks. 
The galaxy was centred on the ACIS-S3 CCD with an Y-offset of about 1$'$
from the nominal pointing of the chip to better fit in the
$\sim$~8$'\times$8$'$ field. 
The S3 chip (back illuminated) was chosen because of the higher sensitivity in
the soft X-ray band. 
The on-axis {\it Chandra} spatial resolution is $\sim$ 1$''$ corresponding to
15 pc at NGC~2403 (we adopted the distance of 3.18 Mpc from
\citet{mad91}). 
The orientation of the telescope was chosen in such a way that the diagonal of
the chip was parallel to the major axis of the galaxy.
The data were reduced and analyzed with the packages CIAO (version 2.1) and
XSPEC (version 11.0.1). 
We removed high background time by excluding events exceeding $\pm$~3~$\sigma$ 
the mean chip count rate. 
The resulting exposure time of the observation was 34.9 ks. 
The energy range was restricted to 0.4$-$10 keV. 

Figure~\ref{fig2} shows the optical (DSS) and $Chandra$ images of NGC~2403.
In the soft (0.4$-$1 keV) X-ray image (upper right panel) an adaptive
smoothing 
was applied to the original data after the background subtraction. 
The background was subtracted using deep blank-field datasets 
with the
same focal plane temperature as NGC~2403 observations ($-$120~$^{\circ}$C).
The bottom panels of Figure~\ref{fig2} show two 5$''$ smoothed S3 images in the
soft (left, 0.4$-$1 keV) and hard (right, 1$-$10 keV) bands without background
subtraction.
These images show numerous discrete X-ray sources located in the bright part
of the stellar disk of NGC~2403.
The adaptively smoothed image also shows the presence of diffuse emission in
the soft X-ray band.

\subsection{Discrete sources}

Discrete X-ray sources were detected in NGC~2403 using CIAO {\it wavdetect}
on the full resolution S3 image.
We used a threshold of 10$^{-6}$ (4.7 $\sigma$ level, or less than 0.25
false detections per chip).
We looked for sources of different sizes, from 1$''$ to 16$''$, to
detect both point and extended sources and in different energy bands to detect 
both soft and hard sources.
A total of 41 sources were found in the field of the S3 CCD.
Assuming a standard (kT $\sim$ 5 keV)
spectrum, the faintest source has an X-ray flux of $\approx$~10$^{-15}$erg 
cm$^{-2}$ s$^{-1}$ or a luminosity of $\approx$~10$^{36}$ erg s$^{-1}$ at the 
distance of NGC~2403.

Some of these sources are probably background sources unrelated to NGC 2403.
On the basis of the ROSAT Log($N$)$-$Log($S$) \citep{has98},
at the above flux limit and considering the shadowing by the galaxy (with a
typical H~{\small I} column density of 0.5$-$1 $\times$ 10$^{21}$ cm$^{-2}$),
we estimate about 5$-$10 background sources in our S3 chip.

The majority of the detected sources are unresolved and show emission in the
hard band.
Those for which a spectral analysis was possible show binary-type spectra. 
Three sources are extremely bright with observed fluxes (0.4$-$10 keV) 
of 4.3 $\times$ 10$^{-13}$, 6.2 $\times$ 10$^{-13}$, and
1.1 $\times$ 10$^{-12}$ erg~s$^{-1}$~cm$^{-2}$.
If they belong to the galaxy their luminosities are
5.2 $\times$ 10$^{38}$, 7.1 $\times$ 10$^{38}$, and 1.3 $\times$ 10$^{39}$
erg~s$^{-1}$ (see also \citet{kot00}).
These sources have luminosities and spectra typical of the class of the Ultra 
Luminous X-ray Sources found in several nearby galaxies \citep{kin01}.
Most of the X-ray sources of NGC~2403 are inside the stellar disk, tracing  
the spiral pattern of the galaxy.
This suggests a relation with the stellar population I 
as found for other late-type spiral galaxies (e.g.\ M~33, \citet{lon96}). 
Three sources coincide with bright H~{\small II} regions and are clearly 
resolved in the {\it Chandra} images (Figure~\ref{fig2}).

\section{Diffuse emission}

Point sources (except the 3 H~{\small II} regions) were removed from the full
resolution {\it Chandra} data of NGC~2403 by masking elliptical areas around
each source.
Each elliptical area found by {\it wavdetect} (3 $\sigma$ of the PSF) was
visually inspected in order to enlarge, if necessary, the masking area.
Pixels inside the ellipses were filled by local background values (Poisson method
in $dmfilth$).
Figure~\ref{fig3} (left) shows the exposure-corrected 
X-ray {\it Chandra} image in the soft band (0.4$-$1
keV) after the subtraction of the point sources (H~{\small II} regions excluded) 
and smoothed to a resolution of 15$''$. 
The contribution to the diffuse emission from sources under the detection limit
can be estimated from the luminosity function.
We obtained the luminosity function from the log($N$)$-$log($S$) calculated
for the 38 point sources and assuming a standard spectrum for X-ray binary
with $\Gamma$ = 1.6 \citep{ten01}.
The best power-law fit for luminosities above 2 $\times$ 10$^{36}$ erg s$^{-1}$ 
is $N(>L)=38.04~ L_{36}^{-0.61}$ where $N(>L)$
is the number of the sources above a given luminosity limit
and $L_{36}$ is the luminosity in units of 10$^{36}$ erg s$^{-1}$. 
From this function we estimate the contribution of sources under the detection
limit to be lower than about 5\%.

Therefore the X-ray emission in Figure~\ref{fig3} is most likely 
produced by truly hot interstellar gas.
The right panel of Figure~\ref{fig3} shows, for comparison, 
an H${\alpha}$ line image of NGC~2403 taken from the archive of the CFH
Telescope.
The distributions of H$\alpha$ and X-ray emission are very similar and the
main features agree in position very well. 
The hot gas seems to follow the spiral structure suggesting that most of 
this emission comes from the disk of NGC~2403 and is related to the star
formation (e.g. see \citet{tru91}).
The emission from the halo, that is expected to be uniform, is not dominant here.

The spectrum of the diffuse emission was extracted with the
software packages {\it calcrmf} and {\it calcarf} provided by A.
Vikhlinin\footnote{http://asc.harvard.edu/cont-soft/software/calcrmf.1.08.html}.
The response matrices were determined from the FITS embedded
function response file FP$-$120 for the operating temperature of $-$120~$^{\circ}$C. 
The spectrum was obtained from an elliptical region of diameter 8$'$ with the 
same position angle (p.a.=124$^{\circ}$) as the disk of NGC~2403.
In order to perform a spectral analysis a careful estimate of the background is needed.
We extracted two background spectra:
the first was extracted from the same elliptical region as that of the
galaxy on the blank field data;
the second was obtained from an elliptical region (about 4$'$ diameter)
in the upper left corner of our S3 chip.
The two backgrounds are similar in the soft (0.4$-$1 keV) band
(1.0 $\times$ 10$^{-7}$ and 1.2 $\times$ 10$^{-7}$ cts s$^{-1}$ pix$^{-1}$
for the first and the second respectively) while above 1 keV the
background of our S3 field is higher (by a factor 2)
than the one of the blank field.
This difference is not caused by background variations within the chip.
We tested this by extracting backgrounds from different regions across the
blank field chip and finding no significant variations.
The contamination of our S3 background by the galaxy is also excluded because
the differences show up in the spectral region (1$-$2 keV) in which the
galactic emission is much weaker.
Besides, we also extracted a third background from an elliptical region on the S1
chip (back illuminated) and found it very similar to the S3 one.
Therefore there is probably a slighly intrinsic difference between the background
during our observations and that on the blank field above 1 keV.
This could be caused by a local anisotropy in the pointing direction or by a
different instrumental noise.
In the following spectral analysis we preferred to use the background obtained 
from our S3 chip.

Figure~\ref{fig4} shows the spectrum of the X-ray emission of NGC~2403 after
the subtraction of point sources in the band 0.4$-$2 keV.
A fit with a Mekal thermal plasma model \citep{mew86} with one component gives
a temperature of kT $\simeq$ 0.25$\pm$0.1 keV (T $\simeq$ 2.9 $\times$ 10$^6$ K) with a
${\chi_r}^2$=1.72.
Better one-component fits (${\chi_r}^2~\simeq$ 1.5)
are obtained with a 0.1 solar abundance or an under-abundance
of alpha elements (O, Mg, Si, Ca) of about 0.3 solar.
A significantly better fit (${\chi_r}^2 \sim$ 1), solid line in Figure~\ref{fig4},
is obtained with two components at temperatures kT$_1\simeq$ 0.18$\pm$0.03 keV
(T $\simeq$ 2.1 $\times$ 10$^6$ K) and kT$_2\simeq$ 0.73$\pm$0.07 keV (T $\simeq$ 8.4 $\times$
10$^6$ K) and absorption column density fixed to the galactic value of
4.1 $\times$ 10$^{20}$ cm$^{-2}$.
Given the poor statistics, the metal abundances are substantially 
unconstrained giving acceptable results 
for values from 0.1 to 1 solar (see also \citet{dah00}).
In the fit in Figure~\ref{fig4} the abundances are fixed to 0.3 solar.
The ``emission line'' at about 1.8 keV is due to a known strong instrumental background 
feature\footnote{http://cxc.harvard.edu/contrib/maxim/bg/index.html\#spec},
not completely subtracted. 

The total observed flux of the diffuse emission (0.4$-$2 keV) is 1.77 $\times$
10$^{-13}$ erg cm$^{-2}$ s$^{-1}$ 
(total luminosity of 2.14 $\times$ 10$^{38}$
erg s$^{-1}$). 
This value corresponds to about 3\% of the total X-ray luminosity from
NGC~2403 and 
is similar to that of 1 $\times$ 10$^{38}$ erg s$^{-1}$ found with ROSAT
for M~33 \citep{lon96}. 
The total de-absorbed luminosity is L$_X$=2.66 $\times$ 10$^{38}$ erg s$^{-1}$.
The relative contributions of the soft and hard components to the total
luminosity are about equal
(L$_{soft}$/L$_{hard}~\sim$ 1 
corresponding to a ratio of the counts rates of
cts$_{soft}$/cts$_{hard}~\sim~$1.4).

\section{Discussion and conclusions}

We have detected diffuse soft X-ray emission from the disk of the spiral
galaxy NGC~2403. 
The X-ray emission is strongly associated with the H~{\small II} regions 
and is most likely produced by hot gas from stellar winds and
expanding supernova shells within the disk of the galaxy. 
If we assume a rate of supernova explosions of 0.01 events yr$^{-1}$ for NGC~2403
\citep{mat97} and a mean energy release per supernova of 10$^{51}$
erg we estimate that the total amount of energy released by supernovae is
$\sim$~3 $\times$ 10$^{41}$ erg s$^{-1}$.
This is much larger than the observed X-ray luminosity (2.1
$\times$ 10$^{38}$ erg s$^{-1}$).
Thus the hot ISM of NGC~2403 can be easily produced by energy input from
star formation activity and a heating efficiency as low as 0.001.

\begin{figure}
\plottwo{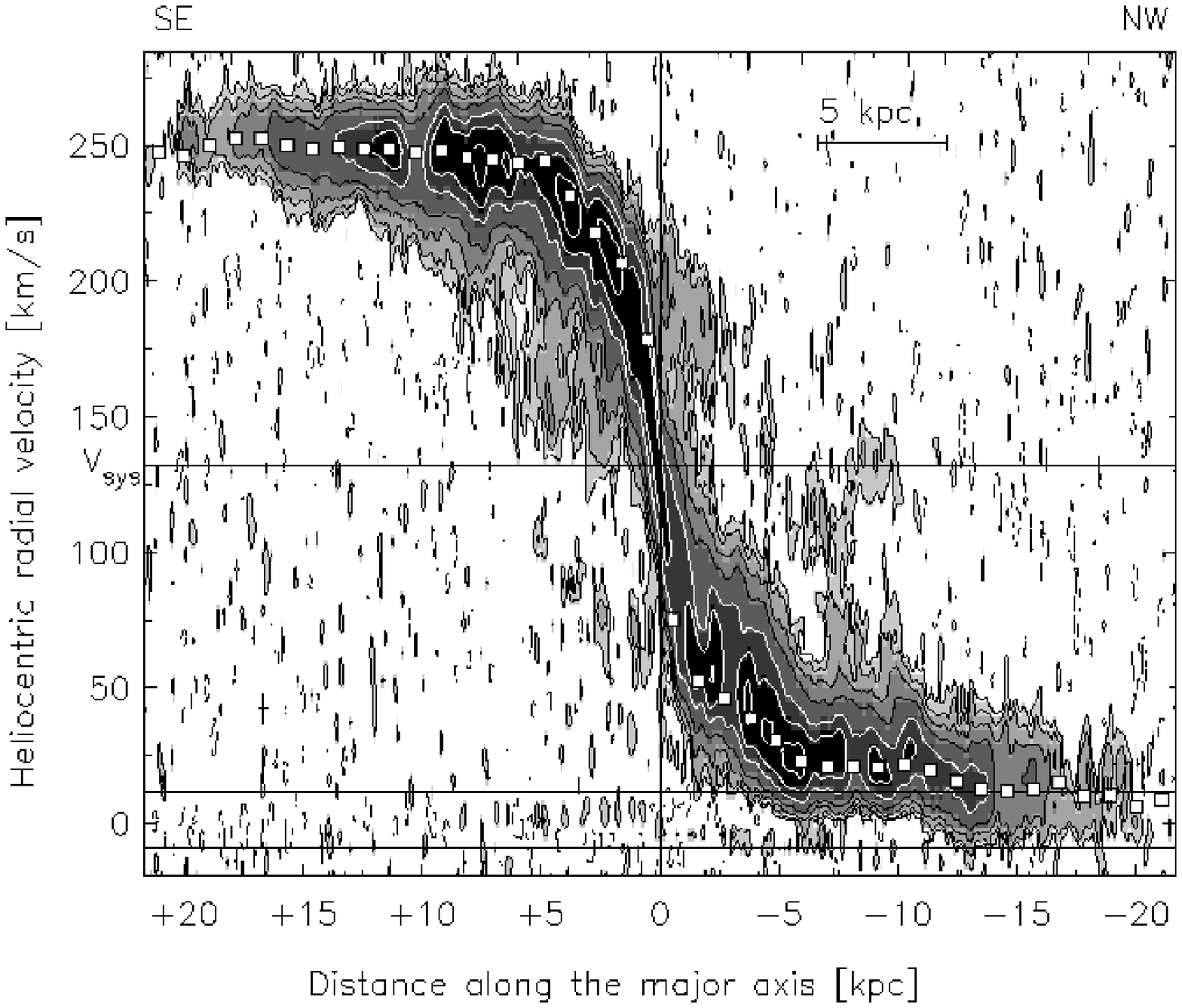}{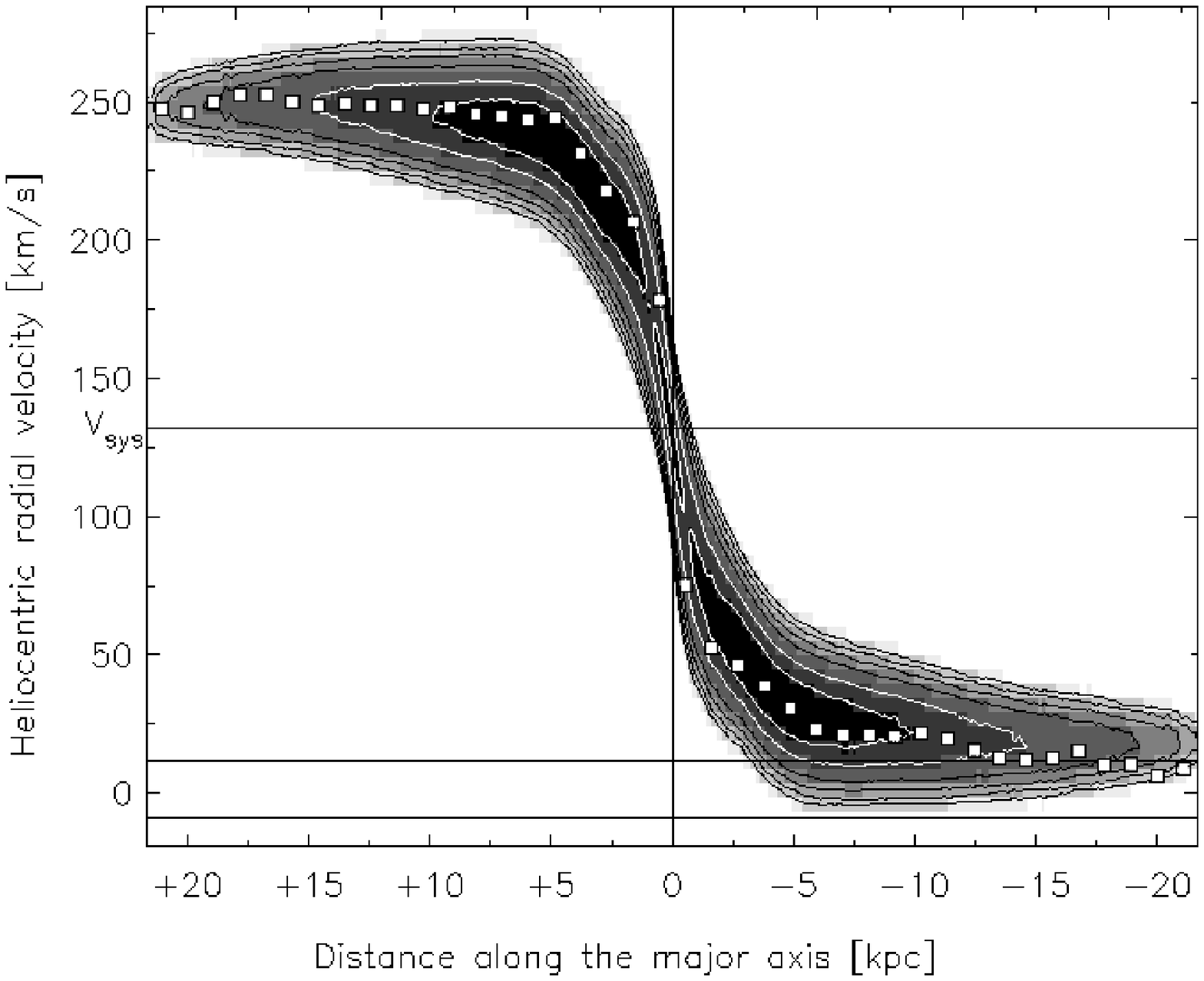}
\caption{H~{\small I} position-velocity diagram along the major axis
(p.a.=124$^{\circ}$, V$_{sys}$=133 km~s$^{-1}$) of NGC~2403.
The beam size is $\sim$~15$''$, the velocity resolution is 10.3 km~s$^{-1}$.
The contours are $-$0.4, 0.4, 1, 2, 4, 10, 20, 40 mJy/beam,
the r.m.s.\ noise is 0.17 mJy/beam.
The white squares show the (projected) rotation curve.
The right panel shows a model for a thin H~{\small I} disk with Gaussian velocity
dispersion.
All the H~{\small I} seen at anomalous velocities (not-reproduced by the model) is
what we call ``anomalous gas''. 
\label{fig1}}
\end{figure}

\begin{figure}
\plotone{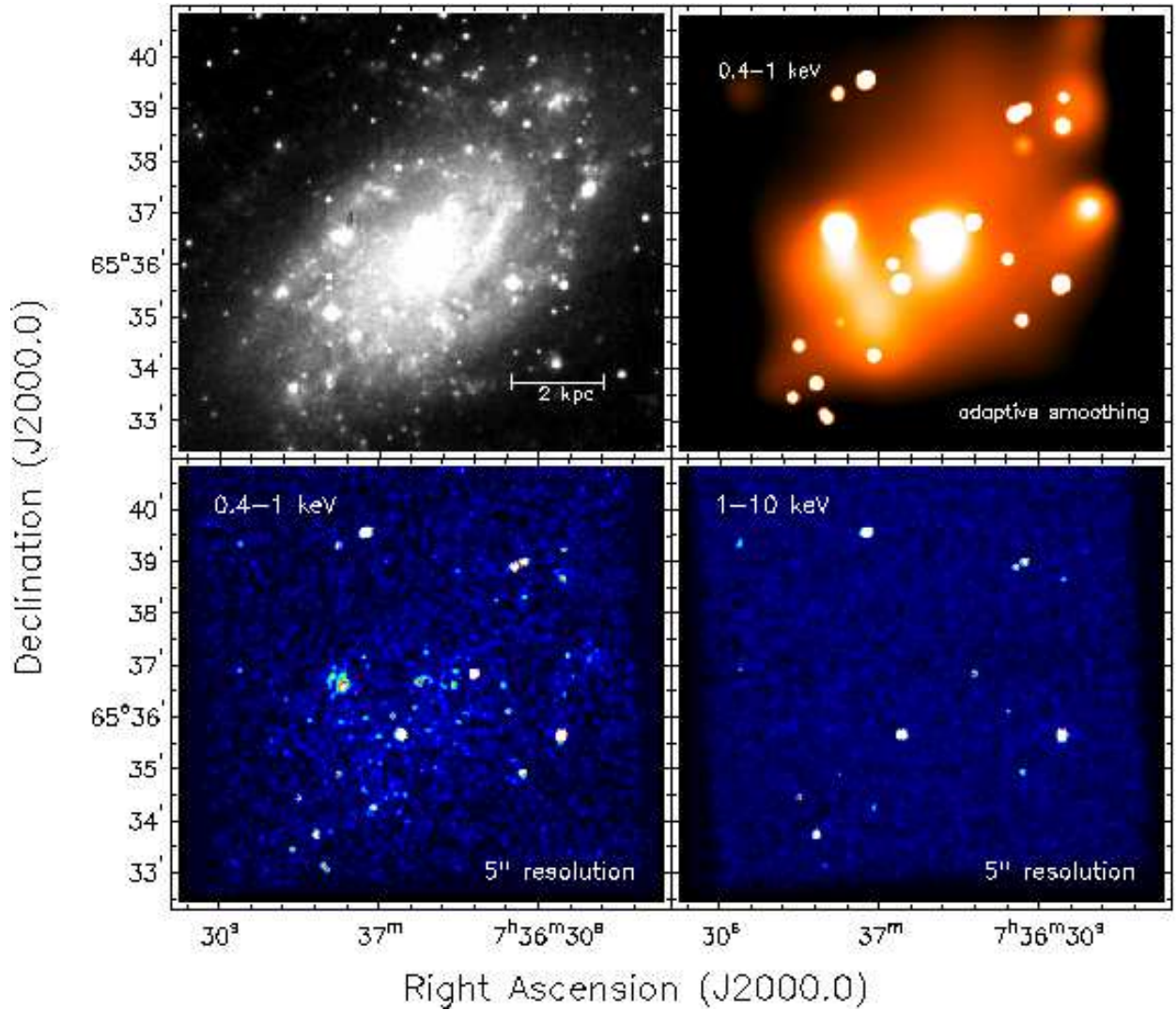}
\caption{Optical DSS (upper left panel) and X-ray S3 {\it Chandra} 
images of NGC~2403 (all on the same scale).
The X-ray image in the upper right panel is the result of an adaptive smoothing
in the soft energy band (0.4$-$1 keV) after the background subtraction. 
The bottom panels show two 5$''$ smoothed S3 images in the soft (left, 0.4$-$1 keV)
and hard (right, 1$-$10 keV) bands. 
\label{fig2}}
\end{figure}

\begin{figure}
\plotone{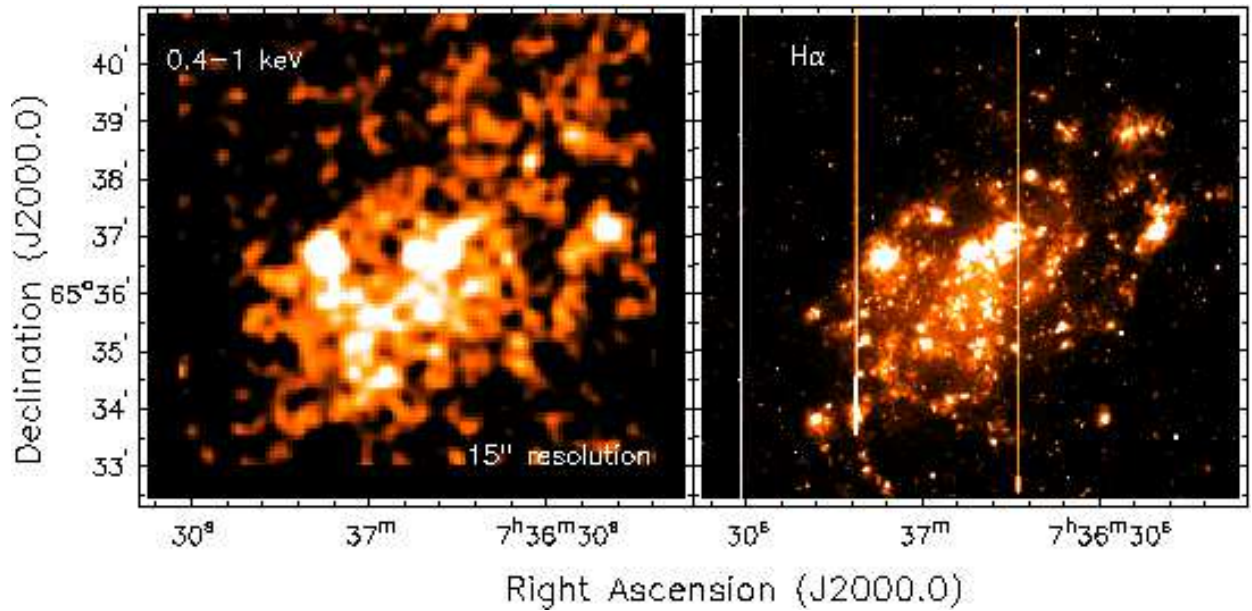}
\caption{Diffuse X-ray emission (0.4$-$1 keV) after the
subtraction of point sources (left panel) and H$\alpha$ image of NGC~2403
(right panel).
The two images are on the same scale.
The resolution of the X-ray image is 15$''$.
The X-ray grayscale ranges from about 3 to 40 in units of r.m.s.\ 
noise above the background (1~$\sigma$ is 6.8 $\times$ 10$^{-7}$ cts s$^{-1}$
arcsec$^{-2}$). 
Note the great similarity between the distribution of hot and ionized gas. 
The brighter H~{\small II} regions have X-ray counterparts and the hot gas seems to
roughly follow the spiral arms of the galaxy. 
\label{fig3}}
\end{figure}

\begin{figure}
\plotone{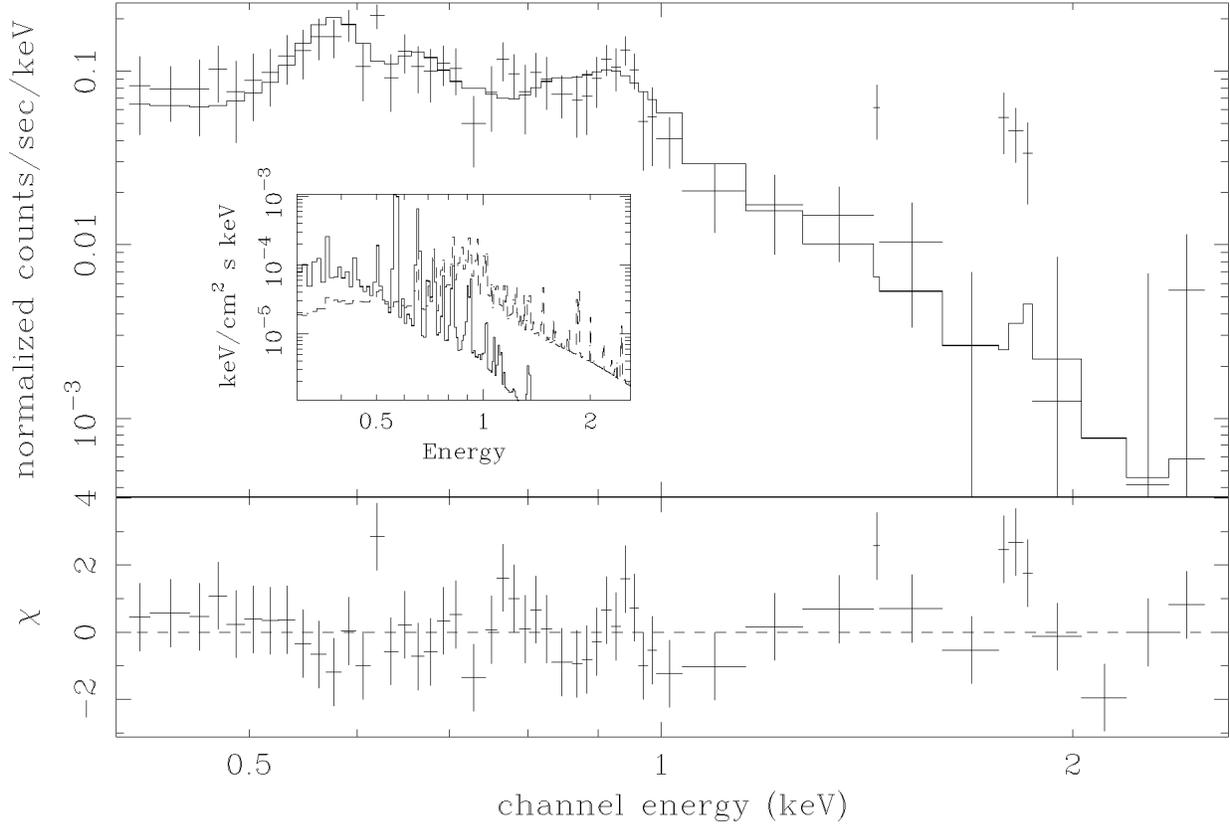}
\caption{
X-ray spectrum of the diffuse emission of NGC~2403 extracted from an
elliptical region of diameter $\sim$ 8$'$ centred on the galactic
centre and with the
same position angle (p.a.=124$^{\circ}$) as the galaxy.
The continuous line is a fit with a two component Mekal model with
temperatures: kT$_1$=0.18 keV, kT$_2$=0.73 keV.
In the inserted small diagram the relative contributions of the two components
are displayed.
In the band 0.4$-$2 keV the two components equally contribute to the X-ray 
luminosity.
In the range 0.5$-$0.8 keV the soft component dominates because of the 
strong emission lines.
\label{fig4}}
\end{figure}

One interesting question is whether the hot gas in NGC~2403 is actually bound
to the galaxy or not.
The escape velocity from a galaxy is usually calculated as
$v_e(r)=\sqrt{2|\phi(r)|}$ where $\phi(r)$ is the potential at radius $r$.
Assuming a mass model (disk $+$ isothermal dark matter halo) for NGC~2403
\citep{fra02} we get an escape velocity of 300$-$350 km s$^{-1}$ 
in the central 4 kpc.
The thermal velocity of the hot gas is likely to be between the adiabatic
sound speed $v_{so}=\sqrt{\frac{5 kT}{3 \mu m_H}}$ and the post-shock 
speed $v_{sh}=\sqrt{\frac{16 kT}{3 \mu m_H}}$.
If we take the temperature of the one-component Mekal fit kT $\simeq$ 0.25 keV 
(T $\simeq$ 2.9 $\times$ 10$^6$ K) we obtain a thermal velocity for the
hot gas (n$_{He}$/n$_{H}$=0.1) of 170$-$300  km s$^{-1}$,
lower than the escape velocity.
Similar values have been found for other galaxies (e.g. NGC~4631, \citet{wan95} and 
the Milky Way, \citet{kan97}).
The above estimate shows that at least part
of the X-ray emitting gas found in NGC~2403 is likely to be 
bound to the galaxy.
One implication of this result is that this hot gas does not substantially
contribute to the enrichment of the surrounding intergalactic medium.
Furthermore, if this gas can
temporarily escape from the disk into the halo region it will necessarily fall
down onto the disk.
According to a ``galactic fountain'' type of process the hot gas escaping 
from the disk is expected to cool quickly because of adiabatic expansion and in
the final phase it could become observable as 
neutral gas at anomalous velocities.
The anomalous H~{\small I} shown in the p-v diagram of Figure~\ref{fig1} is likely
to represent such a final phase of the galactic fountain,
whereas the hot X-ray emitting gas would pertain to the initial phase of it.

From the above X-ray luminosity and temperature we estimate an electron density and
a mass of the hot gas of 0.15 $\times$ 10$^{-3}$ $h^{-\frac{1}{2}} 
f^{-\frac{1}{2}}$ cm$^{-3}$ and 6.0 $\times$ 10$^{6} h^{\frac{1}{2}} 
f^{-\frac{1}{2}}$ $M_{\odot}$ where $h$ is the scaleheight of the disk in kpc
and $f$ is the filling factor.
The cooling rate of the hot gas can be estimated with the formula \citep{nul84}
$\dot{M}_{cool} \approx \frac{M_{X}}{t_{cool}} = \frac{2 \mu m_H L_X}{3 k T}$ that
for $n_{He}/n_{H}=0.1$ gives $\dot{M}_{cool} \sim 0.01$ $M_{\odot}$ yr$^{-1}$.
If we consider now the mass of the anomalous H~{\small I}, 
a typical size of 10 kpc and the measured infall velocity of 10$-$20 km s$^{-1}$,
we estimate the infalling rate of H~{\small I} to be $\dot{M}_{HI} \sim 
0.3-0.6$ $M_{\odot}$ yr$^{-1}$ larger than the cooling rate of the hot gas.
However such value for cooling rate is probably representative only for the disk
component and would vary substantially once the gas has left the disk itself.
A more useful quantity is the outflowing rate of the hot gas.
Considering the mass and the thermal velocity determined above, and assuming a continous 
cycle beetwen hot and neutral gas, such escaping rate
turns out to be $\dot{M}_{hot} \sim$ 0.1$-$0.2 $M_{\odot}$ yr$^{-1}$, of the same order
as the infalling H~{\small I}.
Therefore it is possible that cooling of the X-ray emitting gas in NGC~2403 and its
motion via a galactic fountain could produce the observed H~{\small I} 
position-velocity pattern. 

Is has been suggested that the anomalous H~{\small I} is common among spiral galaxies
\citep{fra01}.
Similarly, the hot X-ray emitting gas 
is probably present in several spiral galaxies
with a high rate of star formation.
One can argue, therefore, that the hot gas and the anomalous H~{\small I} are, generally,
connected as in NGC~2403.
Finally this discovery of diffuse X-ray emitting gas in NGC~2403 has also interesting 
implications for the study of our Galaxy.
The anomalous H~{\small I} in NGC~2403 is analogous to at least some of the High
Velocity Clouds \citep{wak97} of our Galaxy.
The suggested relation between the neutral and the hot gas in 
NGC~2403 supports the possibility that galactic fountains play an important role
also in the Milky Way.

\begin{acknowledgments}
{We thank J. Bregman and G.G.C. Palumbo for helpful comments and advice.
We are grateful to an anonymous referee for various improvements in the paper.
This work received partial financial support from the Italian Space Agency (ASI).
}
\end{acknowledgments}

\clearpage

\clearpage

\clearpage

\clearpage

\clearpage

\begin{thebibliography}{}
\bibitem[Bregman(1980)]{bre80} Bregman, J.N. 1980, \apj, 236, 577
\bibitem[Bregman \& Houck(1997)]{bre97} Bregman, J.N., Houck, J.C. 1997, \apj, 485, 159
\bibitem[Dahlem et al.(2000)]{dah00} Dahlem, M., Parmar, A., Oosterbroek, T., Orr, A., Weaver, K.A., Heckman, T.M. 2000, \apj, 538, 555
\bibitem[Ehle, Pietsch, \& Beck(1995)]{ehl95} Ehle, M., Pietsch, W., \& Beck R. 1995, \aap, 295, 289
\bibitem[Fabbiano(1989)]{fab89} Fabbiano, G. 1989, \araa, 27, 87 
\bibitem[Fraternali et al.(2001)]{fra01} Fraternali, F., Oosterloo T.,
Sancisi, R., Van Moorsel, G. 2001, \apj, 562, L47
\bibitem[Fraternali et al.(2002)]{fra02} Fraternali, F., Van Moorsel, G., Sancisi, R., Oosterloo T. 2002, \aj, 123, 3124
\bibitem[Hasinger et al.(1998)]{has98} Hasinger, G., Burg, R., Giacconi, R., Schmidt, M., Trumper, J., Zamorani, G. 1998, A\&A, 329, 482\bibitem[King et al.(2001)]{kin01} King, A.R., Davies, M.B., Ward, M.J., Fabbiano, G., Elvis, M. 2001 \apj, 552, L109 
\bibitem[Kaneda et al.(1997)]{kan97} Kaneda, H., Makishima, K., Yamauchi, S., Koyama, K., Matsuzaki, K., Yamasaki, N.Y. 1997, \apj, 491, 638
\bibitem[Kotoku et al.(2000)]{kot00} Kotoku, J., Mizuno, T., Kubota, A., Makishima, K. 2000, PASJ, 52.1081
\bibitem[Long et al.(1996)]{lon96} Long, K.S, Charles, P.A., Blair, W.P., Gordon, S.M. 1996, \apj, 466, 750
\bibitem[Madore \& Friedman(1991)]{mad91} Madore, B.F., Freedman, W.L. 1991, \pasp, 103, 933
\bibitem[Matonick et al.(1997)]{mat97} Matonick, D.M., Fesen, R.A., Blair,
W.P., Long, K.S. 1997 \apjs, 113, 333
\bibitem[Mewe, Lemen, \& Van den Oord(1986)]{mew86} Mewe, R., Lemen, J.R., \& Van den Oord, G.H.J. 1986, \aaps, 65, 511
\bibitem[Nulsen, Stewart \& Fabian(1984)]{nul84} Nulsen, P.E.J., Stewart, G.C., Fabian, A.C. 1984, \mnras, 208, 185 
\bibitem[Shapiro \& Field(1976)]{sha76} Shapiro, P.R., \& Field, G.B. 1976, \apj, 205, 762
\bibitem[Snowden \& Pietsch(1995)]{sno95} Snowden, S.L., Pietsch, W. 1995, \apj, 452, 627
\bibitem[Swaters, Sancisi, \& Van der Hulst(1997)]{swa97} Swaters, R.A., Sancisi, R., \& van der Hulst, J.M. 1997, \apj, 491, 140
\bibitem[Tennant et al.(2001)]{ten01} Tennant, A.F., Wu, K., Ghosh, K.K., Kolodziejczak, J.J., Swartz, D.A. 2001, \apj, 549, L43
\bibitem[Truemper et al.(1991)]{tru91} Truemper, J., Hasinger, G., Aschenbach, B., Braeuninger, H., Briel, U.G. 1991, \nat, 349, 579
\bibitem[Valinia et al.(1998)]{val98} Valinia, A., Henriksen, M. J., Mushotzky, R. F.,
 Loewenstein, M., Roettiger, K., Madejski, G. M. 1998, \apj, 505, 134
\bibitem[Wakker \& Van Woerden(1997)]{wak97} Wakker, B.P., \& van Woerden H. 1997, \araa, 35, 217
\bibitem[Wang et al.(1995)]{wan95} Wang, Q.D., Walterbos, R.A.M., Steakley, M.F., Norman, C.A., Braun, R. 1995 \apj, 439, 176 
\bibitem[Wang et al.(2001)]{wan01} Wang, Q.D., Immler, S., Walterbos, R., Lauroesch, J.T., Breitschwerdt, D. 2001, \apj, 555, L99 

\end{thebibliography}
\end{document}